\documentstyle[prl,preprint,aps]{revtex}
\begin{document}
\draft
\title{Study of composite fermions: Beyond few particle systems}
\author{J.K. Jain$^{1,2}$ and R.K. Kamilla$^{1}$}
\address{$^{1}$Department of Physics, State University of New York
at Stony Brook, New York 11794-3800\\
$^{2}$Theoretical Physics Group, Tata Institute of Fundamental Research,
Homi Bhabha Road, Mumbai, 400005, India}
\date{\today}
\maketitle
\begin{abstract}

We construct a new representation of composite fermion wave functions
in the lowest Landau level which enables Monte Carlo computations 
at arbitrary filling factors for a fairly large number of 
composite fermions, thus clearing the way toward a 
more detailed quantitative investigation of the fractional quantum 
Hall effect. As an illustrative application, thermodynamic 
estimates for the transport gaps of several spin polarized 
incompressible states have been obtained. 

\end{abstract}


Interacting electrons confined to two dimensions and subjected to a strong
magnetic field exhibit spectacular phenomena, e.g., the fractional
quantum Hall effect (FQHE) \cite {Tsui}. 
A rather simple and succinct qualitative explanation 
of these phenomena is given in terms of objects called composite fermions 
 \cite {Jain}, 
which are electrons bound to an even number of vortices of the many
body wave function. The FQHE is a manifestation of the Landau level 
(LL) structure of 
composite fermions, and several recent experiments \cite {semi}
have produced striking 
additional evidence for composite fermions by detecting their semiclassical 
cyclotron orbits in the vicinity of the half filled LL \cite {HLR}.

A detailed {\em quantitative} description of the FQHE and related phenomena
is less than satisfactory, however.  It relies largely 
on exact diagonalization studies \cite {Gaps,damb},
which have played an extremely useful role in testing 
and confirming various theoretical postulates, but 
whose predictive power 
is rather limited as they typically deal with systems
containing fewer than 10-12 electrons. Not only is this insufficient
for making reliable thermodynamic estimates in most cases, 
a large region of filling factors is totally inaccessible in such small 
systems. To see this, take the 
conventional spherical geometry. Here, the $n/(2n+1)$ FQHE 
state requires at least $n^2$ electrons (which is the minimum 
number of composite fermions needed to fill $n$ LLs). As a result, only 
two data points are available for 3/7 ($N=9$ and 12) in present 
exact diagonalization studies, and 4/9,
5/11, {\em etc}. do not show up at all.
The situation is worse for the $n/(4n+1)$ sequence -- even 3/13 is out
of reach. Due to an exponential increase with the number of electrons 
in the computer time and memory requirements, it is
unlikely that exact diagonalization studies will tell us much more in
the future.
 
A more promising approach toward the goal of a better 
quantitative description
of the FQHE is to work with the composite fermion (CF)
wave functions, which 
have been shown to be remarkably close to the exact solutions \cite {Dev} 
and are expected to yield
thermodynamic estimates for various quantities correct to within
a few percent. Due to technical difficulties, however, it 
has not been possible in the past to work with these wave functions 
for more than  $\sim$ 10 composite fermions \cite {exception} 
and, as a result, the CF theory has also in general failed to provide  
quantitative information beyond what was known from exact diagonalization 
studies.
In this Letter, we develop a new representation of the CF wave functions
which enables computations for rather large CF systems. We report 
below results for as many as 40 composite fermions; treatment of 
much bigger systems should be possible in the
future. This constitutes a significant step in our ability to achieve  
detailed quantitative predictions for the FQHE. As a first application  
of this method, we have 
computed the transport gaps for several FQHE states of interest.

The approach described below has been tested in both the disk 
(planar) and spherical geometries; here  we will discuss only the 
latter due to space constraint. This geometry \cite{Wu,Haldane} considers 
$N$ electrons on the surface of a sphere moving under the influence 
of a radial magnetic field. 
The single particle eigenstates are the monopole harmonics $Y_{q,n,m}$
\cite{Wu}, which, after a bit of algebra, can be expressed as 
\begin{equation}
Y_{q,n,m}(\Omega_j)=N_{qnm} (-1)^{q+n-m}2^m e^{iq\phi_j} u_j^{q+m} v_j^{q-m}
\sum_{s=0}^{n}(-1)^s {{n \choose s}} {{ 2q+n \choose q+n-m-s}}
 (v_j^*v_j)^{n-s}(u_j^*u_j)^s\;\;,
\label{mh}
\end{equation}
\begin{equation}
N_{qnm}=2^m\left(\frac{2 l +1}{4}\frac{(l-m)!(l+m)!}{(l-q)!(l+q)!}
\right)
^{1/2}\;\;,
\end{equation}
where $\Omega_j$ represents the angular coordinates $\theta_j$ and $\phi_j$
of the $j$th electron, the total flux through the sphere is  equal to
$2q\phi_0$ ($2q$ is an integer and $\phi_0=hc/e$),
$n=0,1,2,...$ is the LL index, $m=-q-n, -q-n+1, ..., q+n$
labels the degenerate states in the $n$th LL, and $l=q+n$.
It is understood here and below that the binomial coefficient 
${{\gamma \choose \beta}}$ vanishes if either $\beta >\gamma$ or $\beta<0$.
The spinor coordinates are defined as \cite{Haldane} 
$u_j\equiv \cos(\theta_j/2)\exp(-i\phi_j/2)
$ and $v_j\equiv \sin(\theta_j/2)\exp(i\phi_j/2)$ \cite {form}.

The CF theory postulates that the strongly correlated 
liquid of interacting electrons is equivalent to a weakly interacting
gas of composite fermions. The ($2p$) vortices bound to electrons have the 
effect of partly cancelling the external field, and
the composite fermions effectively experience a reduced magnetic field,
given by  $B^*=B-2p\rho\phi_0$, where $B$ is the external field, 
and $\rho$ is the electron density. Equivalently, 
the effective filling factor of composite fermions ($\nu^*$) is 
related to the electron filling factor $\nu$ by $\nu=\nu^*/(2p\nu^*+1)$ 
and, in particular, the FQHE of electrons at $\nu=n/(2pn+1)$ is a 
manifestation of the $\nu^*=n$  integer QHE (IQHE)  of composite fermions. 
The (unprojected) wave function for the CF state at $\nu^*$, $\Phi'_{CF}$,
is related to that of the
electron state at $\nu^*$, $\Phi$, as 
\begin{equation}
\Phi'_{CF}={\cal J}\;\Phi\;\;,
\end{equation}
where the Jastrow factor is given by 
\begin{equation}
{\cal J}\equiv \prod_{j<k}
(u_j v_k-v_j u_k)^{2p} \exp[ip(\phi_{j}+\phi_k)]\;.
\end{equation} 
One problem with $\Phi'_{CF}$ is that it is, in general, not strictly 
in the lowest LL (LLL), and thus
cannot be used as a good quantitative representation of the electron 
state at $\nu$ in the high $B$ limit.
This is remedied by simply throwing away the part of the 
wave function that involves  higher Landau levels and working with 
the remaining wave function ${\cal P} \Phi'_{CF}$, where ${\cal P}$ is 
the LLL projection operator.
It has been demonstrated to be very accurate by
comparison with exact solutions available numerically for small systems 
\cite {Dev}. However, in spite of several attempts, it 
has not been possible to compute with ${\cal P} \Phi'_{CF}$ for  
large CF systems, as  would be necessary for exploiting the full 
quantitative potential of the CF theory.  

Let us outline the basic philosophy behind our approach. There are 
compelling theoretical and experimental reasons to believe that 
$\Phi_{CF}'$ contains the correct physics
and is adiabatically connected to the true electron wave function in the 
lowest LL. Moreover, it has only a small fraction of electrons
outside of the lowest LL \cite {Trivedi}. 
Therefore, the objective is to obtain a LLL wave 
function from $\Phi_{CF}'$ without disturbing it violently.
There is nothing {\em a priori} to choose from between various methods 
that accomplish this goal; of  
course the accuracy of the resulting LLL wave function must be established 
by comparison with exact solutions.
The principal result of this work is 
that we have constructed a LLL wave function for composite fermions
different from ${\cal P}\Phi'_{CF}$ but comparably accurate and,
most importantly, much easier to compute with.  

For simplicity, we will consider below only 
states for which $\Phi$ is a {\em single} Slater determinant,
$\Phi=Det[Y_{i}(\Omega_j)]$;
generalization to the case where $\Phi$
is a linear superposition of Slater determinants is straightforward.
Then, 
\begin{equation}
\Phi'_{CF}={\cal J}\;Det[Y_{i}(\Omega_j)]\;\;.
\label{phi'}
\end{equation}
We write the Jastrow factor as  
\begin{equation}
{\cal J}= \prod_{j\neq k} 
(u_j v_k-v_j u_k)^{p} \exp[i\frac{p}{2}(\phi_{j}+\phi_k)]
= \prod_j J_j^p \;,
\end{equation}
with
\begin{equation}
J_j \equiv  
\prod_{k}^{'} (u_j v_k-v_j u_k) \exp[\frac{i}{2}(\phi_{j}+\phi_k)] \;,
\label{jastrow}
\end{equation}
where the prime denotes the condition $j\neq k$.
$J_j$ has the property that when expanded in 
terms of a linear superposition of single particle eigenstates of the $j$th
electron, all eigenstates correspond to the same monopole strength $q'=
(N-1)/2$ (although with different $m'$).
The Jastrow factor can now be incorporated into the Slater determinant
to give  
\begin{equation}
\Phi'_{CF}=Det[Y_{i}(\Omega_j)J_j^p] \;.
\end{equation}
Now, instead of projecting the {\em determinant} on to the lowest LL, as
had been done previously, we project each matrix element individually
to write
\begin{equation}
\Phi_{CF}=Det[{\cal P}Y_{i}(\Omega_j)J_j^p] \;.
\label{phicf}
\end{equation}
Since the product of two LLL wave functions is also in the lowest LL, 
$\Phi_{CF}$ is guaranteed to be in the lowest LL.
In order to evaluate the projection
${\cal P} Y_{q,n,m}(\Omega_j)\;J_j^p$, we first show that
there exists an operator
${\bf Y}^{q'}_{q,n,m}$ satisfying the property that 
\begin{equation}
{\cal P} Y_{q,n,m} Y_{q',0,m'}= {\bf Y}^{q'}_{q,n,m} Y_{q',0,m'}\;\;, 
\label{op}
\end{equation}
where $Y_{q',0,m'}\sim e^{iq'\phi_j}u_j^{q'+m'}
v_j^{q'-m'}$ is a LLL wave function  at monopole 
strength $q'$. For the present purposes, it 
is important that ${\bf Y}^{q'}_{q,n,m}$ be independent
of $m'$.  To this end, we multiply one of the terms on the right hand
side of Eq.~(\ref{mh})
by the LLL wave function $Y_{q',0,m'}$
and write (with $Q\equiv q+q'$, $M\equiv m+m'$, 
and the subscript $j$ suppressed):
\begin{equation}
e^{iQ\phi} (v^*v)^{n-s}(u^*u)^s u^{Q+M}v^{Q-M}=a_0
e^{iQ\phi}u^{Q+M}v^{Q-M}+ higher\;\; LL \;\; states.
\label{proj}
\end{equation} 
For $|M|>Q$, $a_0$ must vanish, since $|M|\leq Q$ in the lowest LL.
Let us first consider the case $|M|\leq Q$. Multipling both sides
by $e^{-iQ\phi} u^{*Q+M}v^{*Q-M}$ and integrating over the angular 
coordinates gives
\begin{equation}
a_0=\frac{(Q-M+n-s)!(Q+M+s)! (2Q+1)!}{(Q+M)! (Q-M)! (2Q+n+1)!}\;.
\end{equation}
This shows that, apart from an $m'$-independent 
multiplicative constant $(2Q+1)!/(2Q+n+1)!$, the LLL projection of 
the left hand side of Eq.~(\ref{proj}) can 
be accomplished by first bringing all $u^*$ and $v^*$ to the left and
then making the replacement
\begin{equation}
u^*\rightarrow \frac{\partial}{\partial u}\;,
\;v^*\rightarrow \frac{\partial}{\partial v}\;.
\end{equation}
While this prescription is not valid in general for $|M| > Q$, 
it can be shown to produce the correct result  
(i.e., zero)
even for $|M|>Q$ for the LLL projection of states of the form
$Y_{q,n,m} Y_{q',0,m'} $: all terms with non-zero binomial coefficients
have the form $\left(\frac{\partial}{\partial u}\right)^{\alpha} u^{\beta}
\left(\frac{\partial}{\partial v}\right)^{\gamma} v^{\delta}$ 
with either $\alpha>\beta$ (for $M<Q$) or $\gamma>\delta$ (for $M>Q$), 
and consequently vanish.  
Thus,
\begin{eqnarray}
{\bf Y}^{q'}_{q,n,m}
&=& \frac{(2Q+1)!}{(2Q+n+1)!}
N_{qnm} (-1)^{q+n-m}2^m e^{iq\phi_j} \nonumber \\
&&\sum_{s=0}^{n}(-1)^s {{n \choose s}} {{ 2q+n \choose q+n-m-s}}
\left(\frac{\partial}{\partial u}\right)^{s} u^{q+m+s} 
\left(\frac{\partial}{\partial v}\right)^{n-s} v^{q-m+n-s}\;.
\label{left}
\end{eqnarray}
A delightful simplification occurs when one brings all the 
derivatives to the right in Eq.~(\ref{left}) using  
\begin{equation}
\left(\frac{\partial}{\partial v}\right)^{\beta} v^\gamma =
\sum_{\alpha=0}^{\beta} \frac{\beta!}{\alpha !} {{\gamma \choose \beta-
\alpha}} v^{\gamma-\beta+\alpha}
\left(\frac{\partial}{\partial v}\right)^{\alpha}\;,
\end{equation}
and a similar equation for the derivative with respect to $u$ (with 
the summation index $\alpha'$).
The sum over $s$ in Eq.~(\ref{left}) then takes the form
\begin{equation}
\sum_{s=\alpha'}^{n-\alpha} (-1)^{s} {{n-\alpha-\alpha' \choose 
s-\alpha'}}=\sum_{s'=0}^{n-\alpha-\alpha'}(-1)^{\alpha'+s'}
{{n-\alpha-\alpha' \choose s'}} \;,
\end{equation}
which is equal to $(-1)^{\alpha'}(1-1)^{n-\alpha-\alpha'}$ and 
vanishes unless $n=\alpha+\alpha'$.  The only term 
satisfying this condition is one with $\alpha=n-s$
and $\alpha'=s$. Consequently, the derivatives in Eq.~(\ref{left}) 
can be moved to the extreme right and act only on the following 
LLL wave function.  Specializing to 
the case where ${\bf Y}^{q'}_{q,n,m}$ acts on $J_j^p$, we write
\begin{equation}
\left(\frac{\partial}{\partial u_j}\right)^s 
\left(\frac{\partial}{\partial v_j}\right)^{n-s} J_j^p
=J_j^p R_j^{s,n-s}\;\;,
\end{equation}
where
\begin{equation}
R_j^{s,n-s}={\bf U}_j^s {\bf V}_j^{n-s}\;,
\end{equation}
\begin{equation}
{\bf U}_j=J_j^{-p}\frac{\partial}{\partial u_j}J_j^p= p\sum_{k}^{'}\frac{v_k}{
u_j v_k - v_j u_k} + \frac{\partial}{\partial u_j}\;\;,
\end{equation}
\begin{equation}
{\bf V}_j=J_j^{-p}\frac{\partial}{\partial v_j}J_j^p= p\sum_{k}^{'}\frac{
-u_k}{u_j v_k - v_j u_k} + \frac{\partial}{\partial v_j}\;\;.
\end{equation}
Substituting in Eq.~(\ref{phicf}) and factoring out the mini-Jastrow
factors $J_j$ to produce back 
the full Jastrow factor ${\cal J}$ finally gives 
\begin{equation}
\Phi_{CF}={\cal J}\;Det[\tilde{Y}_{i}(\Omega_j)]\;\;,
\end{equation}
which has the same form as the unprojected wave function
$\Phi'_{CF}$ in Eq.~(\ref{phi'}) except that each $Y$ has been replaced by
$\tilde{Y}$, given by 
\begin{eqnarray}
\tilde{Y}_{q,n,m}(\Omega_j) &=& 
N_{qnm} (-1)^{q+n-m}2^m \frac{(2Q+1)!}{(2Q+n+1)!} e^{iq\phi_j}  
u_j^{q+m} v_j^{q-m} \nonumber \\ 
&&\sum_{s=0}^{n}(-1)^s {{n \choose s}} {{ 2q+n \choose q+n-m-s}}
v_j^{n-s} u_j^s R_j^{s,n-s} \;\;.
\end{eqnarray}
A consequence of the LLL projection is that 
changing the coordinates of one particle alters all elements of the
Slater determinant in $\Phi_{CF}$, as $R_j$ depends on all particle 
coordinates. As a result,
certain time saving tricks for updating the Slater determinant 
\cite {Ceperley2} at each step of the Monte Carlo cannot be used
here and the full determinant must be evaluated at each step. This
increases the computation 
time enormously (as compared to Monte Carlo on
$\Phi'_{CF}$), but it is still possible to 
deal with much bigger systems than before. 

Of course, the usefulness of $\Phi_{CF}$ hinges critically on 
how precisely it approximates the true electron state.  
We resort to small systems to answer this question.
Here and below, we will assume fully polarized electrons
confined to the lowest LL.
The Coulomb energy of $\Phi_{CF}$, obtained by 
Monte Carlo, along with the corresponding exact ground state energy
is given in Table I for several incompressible 
states. A comparison establishes the extreme accuracy of 
$\Phi_{CF}$. 

We are now in a position to make detailed quantitative 
predictions for a number of experimentally measurable quantities 
in the FQHE. We consider in this paper the (transport)
gaps of FQHE states belonging to the principal sequence $\nu=n/(2n+1)$. 
The gaps of the 1/3 and 2/5 states have been
known quite well from exact diagonalization studies \cite {Gaps,damb};
the gap of 3/7 is known with a large (40\%) uncertainty \cite {damb};
and no estimates exist for 4/9, 5/11, {\em etc}.
From a different perspective, 
Halperin {\em et al.} \cite {HLR} have suggested, based on 
particle-hole symmetry in the lowest LL and the earlier known
numerical results, that the gaps  of the $n/(2n+1)$ states are 
given by the equation  
\begin{equation}
E_{g}=\frac{C}{|2n+1|}\frac{e^2}{\epsilon \ell}\;\;,
\label{gap}
\end{equation}
with $C\approx 0.31$. 
There is also experimental support for such behavior \cite {Du}.
Equating $E_{g}$ to the effective cyclotron energy 
of composite fermions, $\hbar eB^*/m^* c$, motivated by a mean-field 
description of composite fermions, produces an effective mass
of composite fermions that scales as $m^*\sim \sqrt{B}$ \cite {HLR}. 

The gap of the FQHE state at $\nu=n/(2n+1)$
is equal to the energy required to create a far separated
CF particle-hole pair. The ground state has $n$ filled LL's of composite 
fermions and the excited state is obtained by taking one
composite fermion out from the south pole 
of the $n$th CF-LL and placing it on the north pole of the $(n+1)$th
CF-LL (to maximize the distance between the CF-particle-hole pair).
Since the gap, an $O(1)$ energy, is obtained as the
difference between two large $O(N)$ energies, its accuracy is expected to be
much less than that of either the ground state or the excited state energy. 
To get a feel for how well the CF theory does here, 
Table II quotes the gaps predicted by the CF theory for some of 
the largest systems for which exact gaps are also known;
a comparison indicates that the error in the thermodynamic limit
is expected to be within a few percent, typically much smaller than
the statistical uncertainty of our Monte Carlo.
We proceed to compute the gaps for 2/5, 3/7, 4/9 and 5/11 as a function of 
$N$ \cite {one}. Fig.~1 shows the results for 3/7 and 4/9.
Each error bar signifies one standard 
deviation, as  determined from the average gaps from ten Monte Carlo runs.
The thermodynamic estimates, shown in Fig.~2, have been 
obtained using a chi-square fitting that biases the points by
their error bars. For $\nu=$ 3/7 the gap is consistent with 
the earlier estimate, although with much smaller uncertainty. 
The overall behavior of the gaps is in good agreement with 
Eq.~(\ref{gap}), as seen in Fig.~2.  
It should be noted that the experimental values of the gaps are 
reduced by various unavoidable effects, e.g., finite 
width of the quantum well \cite {Wellwidth}, LL 
mixing \cite {LLmixing} and disorder, and the present estimates 
are actually only the upper bounds. The finite well width 
softens the Coulomb interaction at short distances, which 
is straightforward to incorporate into our Monte
Carlo and will be the subject of a future work. 
LL mixing is harder to deal with.
Fixed phase Monte Carlo \cite {Ceperley} or a variational approach 
considering a linear combination of $\Phi_{CF}'$ and $\Phi_{CF}$ 
\cite {Bonesteel2} may be useful in this context.

In summary, we have developed an approach that permits an investigation 
of large CF systems and thereby makes accessible previously 
unexplored regions of the FQHE. 
We believe that it will prove extremely useful  
toward a better quantitative understanding of the FQHE. 

This work was supported in part by the National Science Foundation
under Grant no. DMR93-18739 and by a fellowship from  
the John Simon Guggenheim Foundation.

\pagebreak

{\bf Table Caption}

Table I. Comparison between the energy per particle of the CF wave function
$\Phi_{CF}$ and the exact Coulomb energy for the ground states at
$\nu= $ 2/5 and 3/7.
The energies are in units of $e^2/\epsilon \ell$,
where $\ell$ is the magnetic length and $\epsilon$ is the background
dielectric constant, and include interaction with the 
uniformly charged positive background. The energy of $\Phi_{CF}$ has
been evaluated by Monte Carlo, with the statistical uncertainty in 
the last two digits shown in brackets. 
The exact results for $N=12$ and $\nu=3/7$ are taken here and 
in Table II from S. He, S.H. Simon
and B.I. Halperin, Phys. Rev. B {\bf 50}, 1823 (1994).

Table II. The gap to create a CF-particle-hole excitation in which 
a composite fermion is removed from the south pole 
of the topmost filled CF-LL and placed on the north pole of the 
lowest empty CF-LL. The exact gaps are also given; the gap at 
$\nu=2/5$ for $N=10$  has been read off
from the spectrum in Ref. \cite {damb}.

{\bf Figure Captions}

Fig. 1. The energy gaps (in units of $e^2/\epsilon \ell$)
at 3/7 and 4/9 for several  
$N$. In order to minimize finite size effects,  the interaction energy
of two appropriately charged pointlike particles
situated at the two poles has been
subtracted from the gaps here (compared to those 
quoted in Tables I and II), following \cite {Gaps,damb}.

Fig. 2. Thermodynamic values of the gaps plotted as a function 
of $1/(2n+1)$. 
The gap for $\nu=1/3$ is taken from Ref.~\cite {Bonesteel}.
The straight line is a chi-square fit, given by 
$E_g=-0.001(5) + 0.320(2) (2n+1)^{-1}$.

\pagebreak

\begin{center}

\begin{tabular}{|c|c|c|c|} \hline
$\nu$ &  $N$   & CF energy & exact energy \\ \hline
\hline
$\frac{2}{5}$   &  6   &    -0.500339(42)  &   -0.5004002 \\ \cline{2-4}
  &  8   &   -0.480216(33) &    -0.4802436 \\ \hline
$\frac{3}{7}$ &   9  &    -0.499138(71) &   -0.4991843 \\ \cline{2-4}
 & 12  &    -0.482507(49)  &   -0.4826388 \\ \hline \hline
\end{tabular}

\vspace{0.5cm}

TABLE I

\vspace{1.0cm}

\begin{tabular}{|c|c|c|c|} \hline
$\nu$ &  $N$   & CF gap& exact gap \\ \hline
\hline
$\frac{2}{5}$  &  6 &    0.07615(61)  &   0.07505 \\ \cline{2-4}
 &   8  &   0.07021(87) &    0.06809 \\ \cline{2-4}
 &  10  &  0.0681(12)  &   0.0673(7) \\ \hline
$\frac{3}{7}$ & 9 &   0.0691(14)  &   0.0681 \\ \cline{2-4}
 & 12  &     0.0518(12)  &   0.0525 \\ \hline
\end{tabular}

\vspace{0.5cm}
TABLE II

\end{center}


\begin{thebibliography}{99}
\bibitem{exception} The only exception is in
the immediate vicinity of $\nu=1/(2n+1)$, where 
it has been possible to study larger CF systems
\cite {Bonesteel,Kamilla}.
\bibitem{form} Using the
identity $u^*u+v^*v=1$, the Eq.~(\ref{mh}) can be cast into
different equivalent forms; for example, the sum over $s$ can be expressed
as a power series entirely in $v^*v$ or $u^*u$. The subsequent
formulas are, however,  
much more complicated for forms other than that in Eq.~(\ref{mh})
for which a number of simplifications are obtained.
\bibitem{one}
In the case of $\nu=1/3$, the present projection 
technique is equivalent to full projection,
investigated earlier \cite {Bonesteel}.
\end{thebibliography}
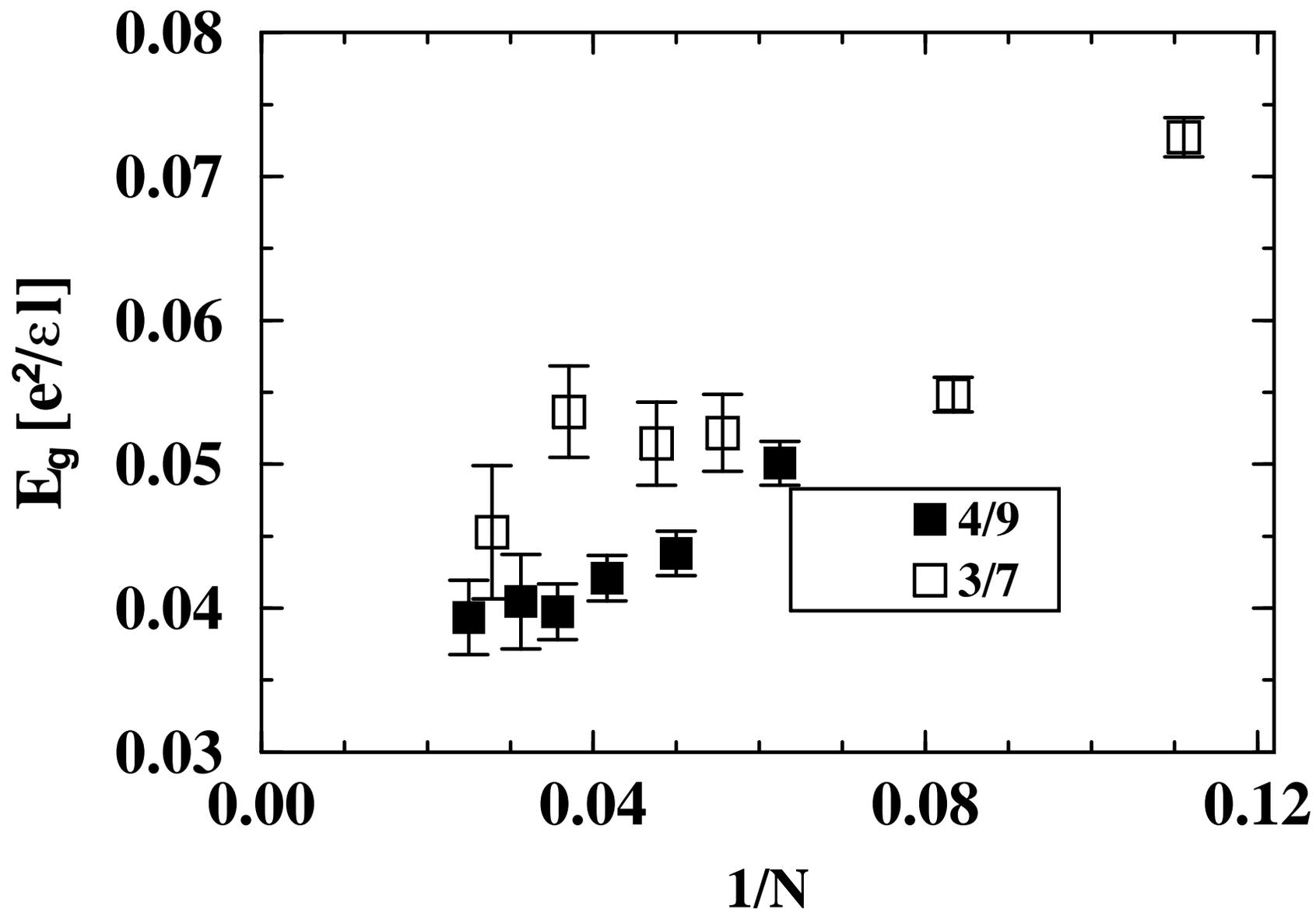
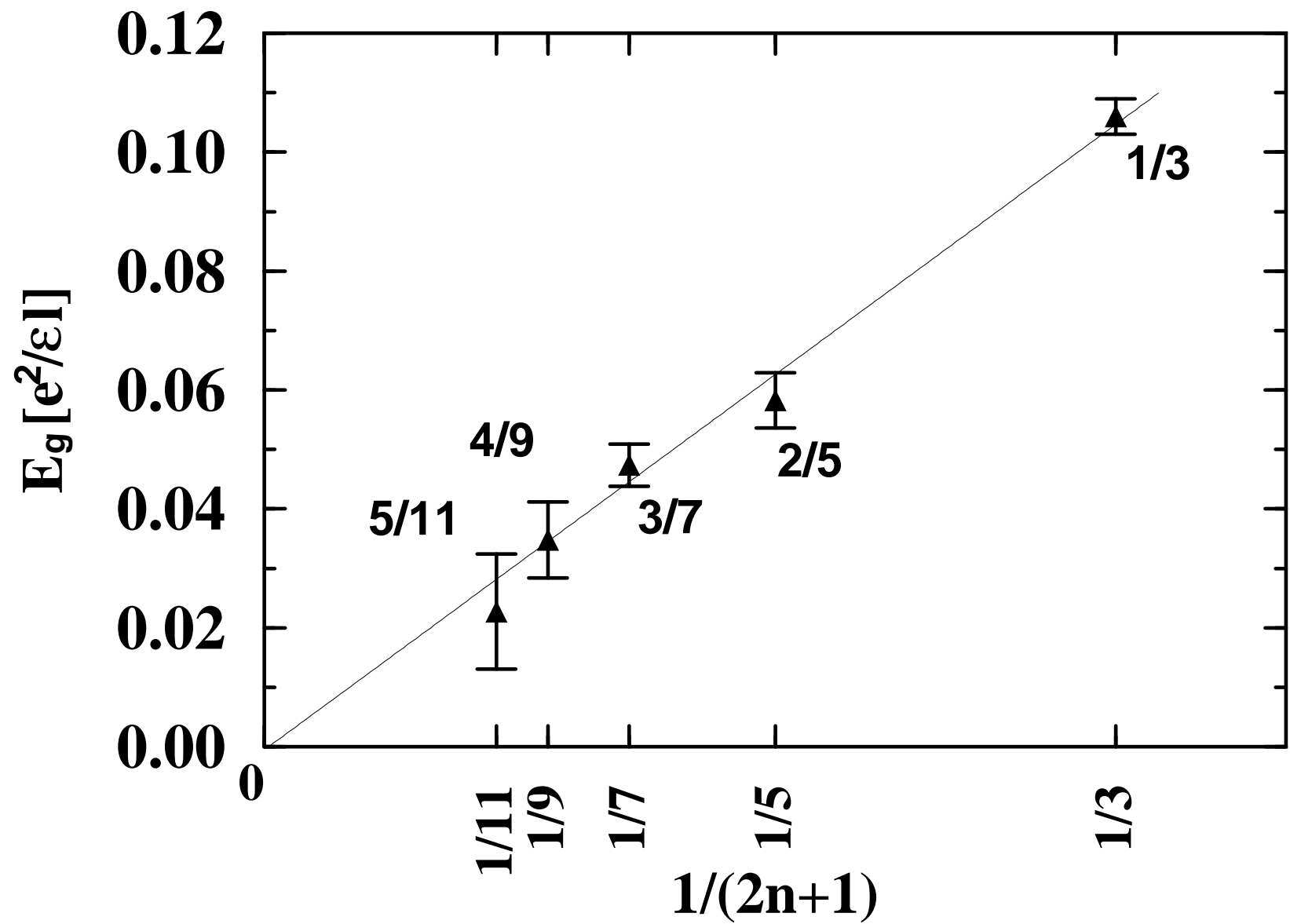
\end{document}